\documentclass[amsmath,amssymb,nofootinbib,showpacs,twocolumn,nofootinbib]{revtex4-1} 
\usepackage{dcolumn}
\usepackage{bm}
\usepackage{graphicx}
\usepackage{epsfig}
\usepackage{color}
\usepackage{wasysym} 
\usepackage{natbib}
\usepackage{twoopt}
\usepackage{dashrule}
\usepackage{MnSymbol} 

\definecolor{mycolor}{rgb}{0.45, 0.31, 0.59}
\usepackage[breaklinks=true,colorlinks=true,linkcolor=mycolor,citecolor=mycolor,urlcolor=mycolor,pdfauthor={Tomassetti},pdftitle={DirectCRMeasurements}]{hyperref}

\usepackage{xspace}
\newcommand{\ApJ}{ApJ}

\newcommand{\PRL}{PRL}
\newcommand{\PRD}{PRD}

\newcommand{\ASR}{ASR}

\newcommand{\etal}{et alii}
\newcommand{\eg}{\textit{e.g.}} 
\newcommand{\ie}{\textit{i.e.}}

\def\citep#1{\cite{#1}}

\begin{document}

\title{Direct Measurements of Galactic Cosmic Rays}
\author{Nicola Tomassetti\,{\color{mycolor}$^{\largestar\,\leftmoon}$}}
\address{{\color{mycolor}$\largestar$}\, Dipartimento di Fisica e Geologia, Universit{\`a} degli Studi di Perugia, Italy\\{\color{mycolor}\leftmoon}\, Istituto Nazionale di Fisica Nucleare – INFN Sezione di Perugia, Italy}
\begin{abstract}  
  This paper reviews recent progress in the field of direct measurements of Galactic cosmic rays.
  High-statistic measurements of cosmic ray energy spectra, chemical and isotopic composition, and the rare antimatter components have been made using large particle physics experiments operating in space. The recent results are discussed in relation to our understanding of the origin of cosmic rays, the open questions, and the challenges for future experiments of direct detection.
\end{abstract}
\maketitle

\section{Introduction}    
\label{Sec::Introduction} 

Cosmic rays (CRs) are the main constituents of the charged radiation in space
and represent one of the most fascinating research subjects in modern astrophysics. 
They are constituted by ionized atoms of all known elements and their isotopes, electrons, and antimatter.
It may be said that high-energy CRs constitute a genuine sample of Galactic matter.
In fact, most of cosmic particles come from outside the solar system and,
in the range from multi-MeV to multi-PeV energies, they are believed to originate in the Milky Way.
More precisely, CRs are called \emph{primary} if they are accelerated by astrophysical sources such as supernova remnants or stellar winds,
while \emph{secondary} CRs are generated by collisions of high-energy nuclei with the interstellar matter.
The most abundant species such as protons and helium are primary CRs.
Also carbon, oxygen, iron and other nuclei synthesized in stars, as well as electrons, are of primary origin.
Rarer species such as lithium, beryllium, boron, fluorine, or titanium (which are not end-products of stellar nucleosynthesis) are secondary particles.
The presence of a secondary component also leads to the conclusion that the abundance observed in CRs in do not reflect the actual source spectrum of these particles. 
Similarly, antimatter particles such as antiprotons and positrons should be mainly of secondary origin.
Whether a small fraction antimatter may originate from primary sources is currently an open question.
The study of secondary elements that are produced by nuclear fragmentation in the interstellar gas can therefore provide key
scientific information about this question.

Direct measurements of CR energy spectrum, composition or arrival directions must be performed at the top of the terrestrial atmosphere by space- or balloons-borne experiments.
The direct detection of CRs can be achieved up to energies where there is sufficient flux, depending on the sensitivity of the experiment.
Since the sixties, several experiments for direct CR detection have been developed onboard stratospheric balloons, satellites, space probes and space stations.
Current measurements extend from the MeV scale up to about 100\,TeV of kinetic energy per nucleon.
Above these energies, the CR flux is inferred up to about 10 EeV of energy with indirect methods, using data from ground-based experiments
after modeling the interaction of CRs with the atmosphere.

In this paper, I will review the recent results from experiments of direct CR detection,
their implications on our understanding on the origin of CR, and the new open challenges for future experiments.
The present paper is based on my review talk given at the 27th European Cosmic Ray Symposium, ECRS, from July 25th to 29th, 2022 in Nijmegen, The Netherlands.
Excellent recent reviews on this topics can be found in Refs.\,\citep{Beatty2022,Boezio2020,PengYuan2019,PicozzaMarcelli2019,Marocchesi2017}.
The rest of this paper is organized as follows.
Section\,\ref{Sec::CosmicRays} gives an overview of the basic phenomenology of Galactic CRs.
Section\,\ref{Sec::GoldenAge} provides a brief summary the main experiments of direct CR detection in the last decades.
Section\,\ref{Sec::Results} presents the recent results in direct CR detection of light, intermediated, and heavy nuclei, as well as light leptons and antiparticles.
In Section\,\ref{Sec::FutureExperiments}, some of the future projects of direct CR detections are discussed.
My conclusions are given in Section\,\ref{Sec::Conclusions}.

\section{The cosmic rays}  
\label{Sec::CosmicRays}    

\begin{figure}[!t]
\centering
\includegraphics[width=0.42\textwidth]{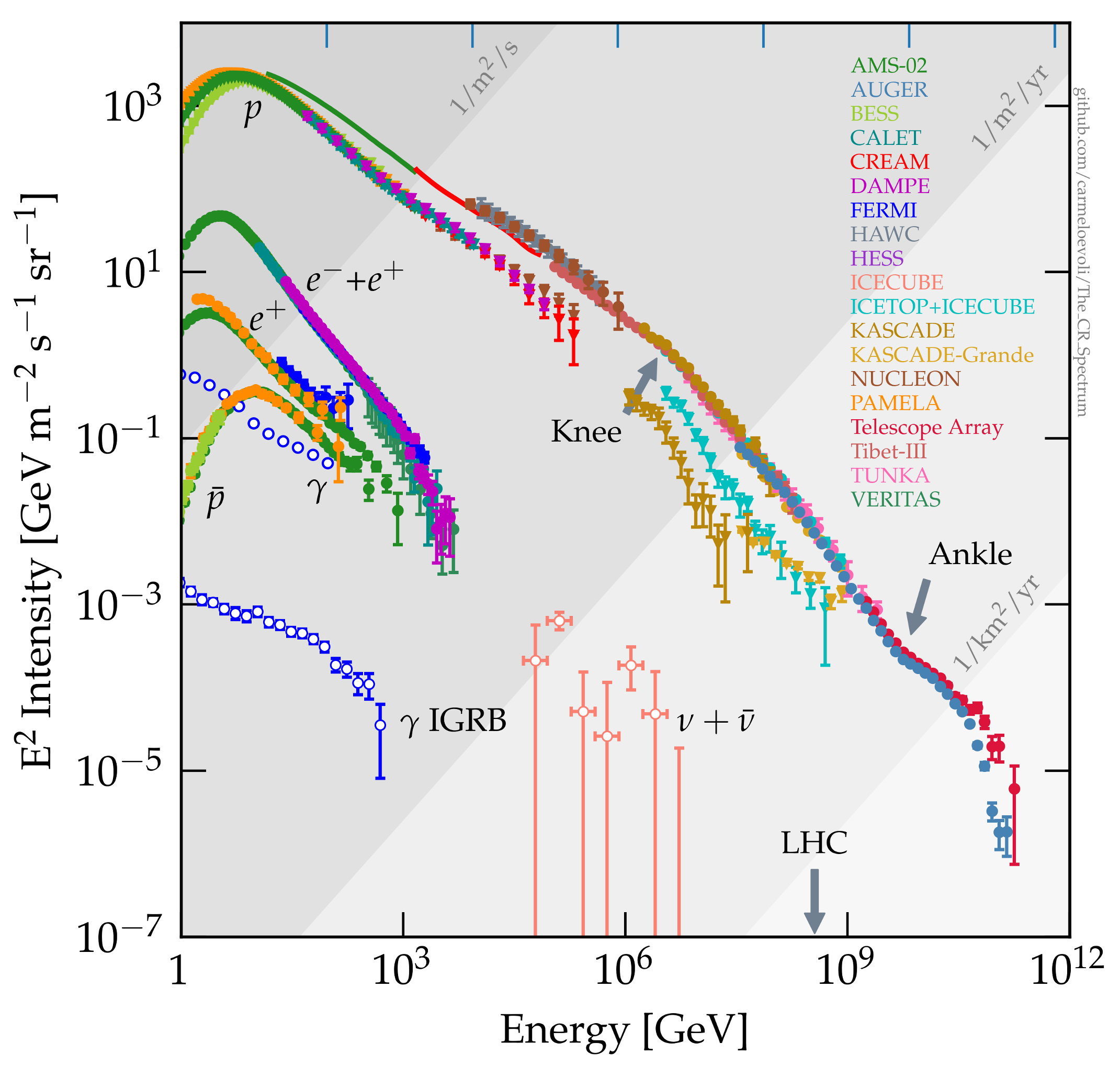}
\caption{\footnotesize%
  Compilation of measurements of the CR flux as function of energy for different charged species, along with gamma-rays and neutrinos.
  Courtesy of \href{https://doi.org/10.5281/zenodo.4309926}{Carmelo Evoli}.
}
\label{Fig::ccCRSpectrum}
\end{figure}

The energy spectrum of CRs follows a steeply falling behavior, nearly power-law behavior, that extends from a few MeV to EeV energies.
Known features appearing in the spectrum are the so-called knee, located at energy of 3 PeV, and the ankle beyond 10 EeV. 
The low-energy part of the spectrum is affected by the solar modulation, that arises from the transport of CRs in the solar wind and the interplanetary magnetic field.
A compilation of measurements of the top-of-atmosphere flux of CRs as function of energy for different species is shown in Fig.\,\ref{Fig::ccCRSpectrum}.
In figure shows the total flux of charged CRs, the so-called ``all-particle'' spectrum, as function of energy.
Other components are the flux of CR protons, electrons, positrons, antiprotons, along with gamma-rays and high-energy neutrinos. 
It can be noted that, in the energy range from GeV to PeV, the CR flux is a rapidly falling function of the particles energy $E$.
This behavior is approximately described by a power-law function $J(E) \cong A(E/E_{0})^{-\gamma}$,
where $A{\approx}1.8\times 10^{4}\,m^{-2}s^{-1}sr^{-1}GeV^{-1}$, $\gamma{\approx}2.7$, and $E_{0}\equiv\,GeV$.

In the seventies, it was also discovered that the flux of secondary CRs is subjected to a rapid energy dependence, with a 
slope $\gamma\gtrsim\,3$, \ie, appreciably higher than that of primary species.
In practice, the secondary to primary flux ratio is observed to decrease with increasing energy.
This feature has been interpreted for long time in terms of an energy dependent confinement time of CRs in the Galaxy.
For the energies concerned in this review, such a confinement time range from $\sim$\,10 kyr to about 10 Myr. 
In the context of modern astrophysical models of CR propagation, the study of energy spectra and charge/mass
composition of primary and secondary CRs can provide key scientific information about their diffusive propagation across magnetic turbulence of the Milky Way.
%
%
In the simplest models of CR propagation, the diffusion of CRs is described by an energy dependent (or rigidity dependent)
coefficient of the type $K(E) \sim E^{\delta}$, with $\delta\approx\,0.3-0.6$.
The confinement time of CRs at energy $E$ is roughly given by $\tau \sim L^{2}/K(E)$, where $L\gtrsim{kpc}$ is the typical size of their confinement region,
and the diffusion coefficient is of the order of $kpc^{2}/Myr$.
Due to diffusive propagation, CRs are expected to be remarkably isotropic at most energies. 
This is indeed the case. For CRs at TeV-scale energy, where the observed level of anisotropy on various angular scales is of about $10^{-3}$, 
presumably due to the global streaming of CRs in the Milky Way through the interstellar magnetic fields.

\begin{figure}[!t]
\centering
\includegraphics[width=0.42\textwidth]{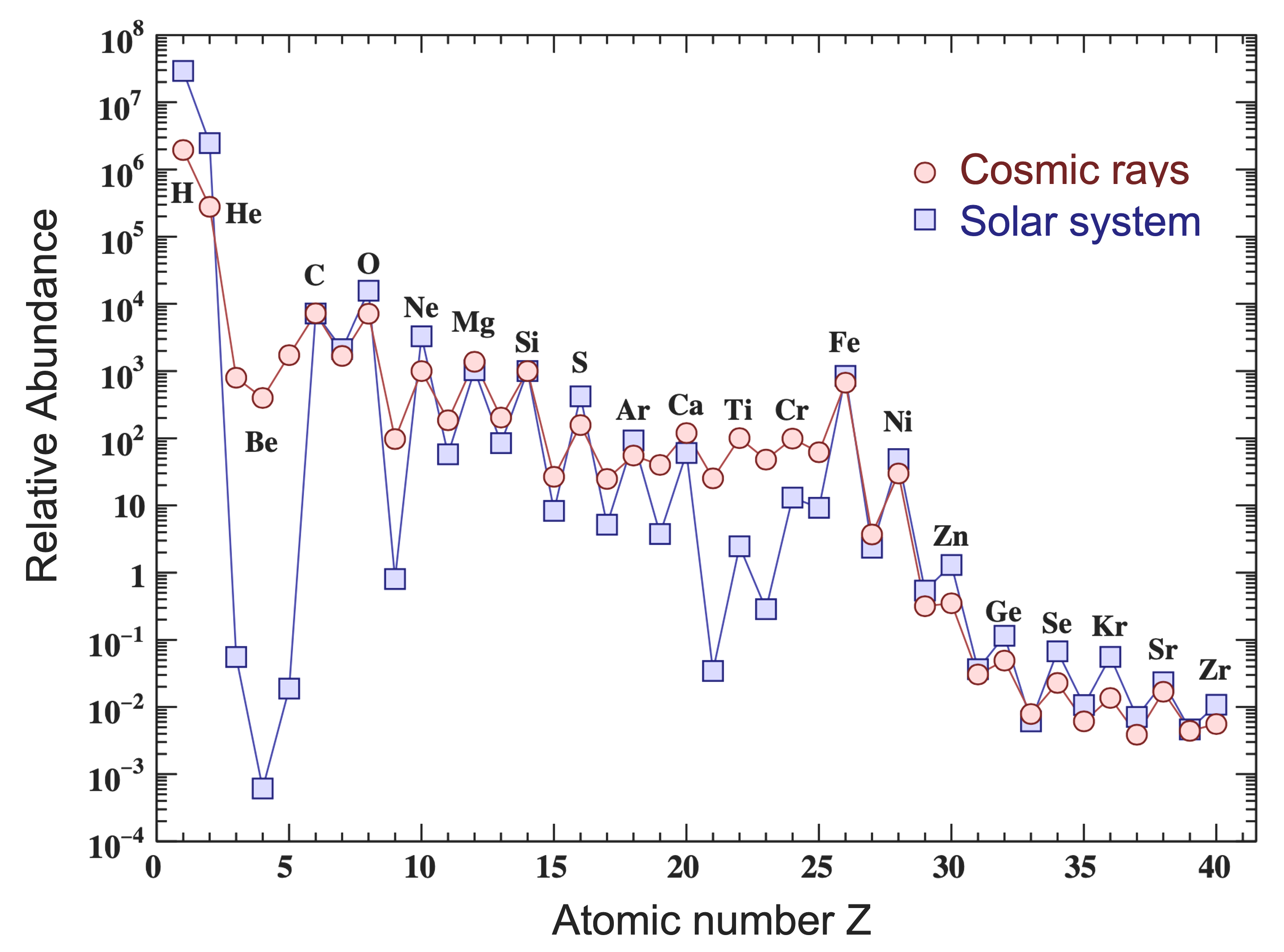}
\caption{\footnotesize%
  Elemental abundances in CRs relative to Si ($\equiv\,10^{3}$) compared to the abundances in the solar system \citep{Beatty2022}. 
  The CR data are from AMS-02 (H,He) \citep{Aguilar2021Report}, ACE/CRIS (Li-Ni) \citep{Lave2013ACECRIS}, and SuperTIGER (Cu-Zr) \citep{Murphy2016SuperTiger}.
}
\label{Fig::ccCRAbundance}
\end{figure}

The relative composition of the various CR elements is shown in Fig.\,\ref{Fig::ccCRAbundance}. In the figure, the CR abundances are also compared with the
abundance of elements in the solar system, which is also well representative of the interstellar medium.
It can be seen that about 88\% of CRs are constituted by ionized hydrogen (protons), and about 9\% are helium nuclei.
Heavier nuclei and electrons contribute to a few percents of the total CR abundance.
From the comparison with solar abundance, secondary CRs contribute to enriched abundances of the rarer elements such as, in particular,
those of the Li-Be-B and group or the sub-Fe group.
A similar over-abundance in CRs is expected for antiparticles such as positrons or antiprotons. 
From the figure, a reduced even-odd effect can also be noticed in CRs.
%
%

In the last decades, an impressive amount of data have been collected on the CR energy spectrum, arrival directions, and chemical composition.
According to the mainstream interpretations of the observation accumulated in a century of observations,
we can formulate the \emph{standard paradigm} of CR physics from three main pillars: 
(i) CRs are interstellar particles accelerated by diffusive shock acceleration processes in supernova remnants;
(ii) they propagate diffusively through interstellar turbulence over an extended magnetic halo, (iii) they interact with the gas nuclei
of the interstellar medium to produce secondary particles.
Such a paradigm has been successful for decades in explaining the observations.
The many concrete models based on these hypotheses make use of various simplifying assumptions regarding sources and diffusion 
properties such as, \eg, homogeneity, isotropy, stationarity, and linearity.
Several longstanding questions about CRs and their sources are still waiting for conclusive answers, such as:
What’s the CR composition in their sources?
Are different CR species accelerated from the same sources?
Which sources contributes to CRs and at which energies?
How the acceleration mechanism works? How CR propagation is related to the Galactic turbulence?
Precise measurements of the energy spectra and charge/mass composition of CRs become essential tools
to scrutinize the paradigm and improved the various models based on the above hypotheses.

\section{A golden age of new CR measurements} 
\label{Sec::GoldenAge}                        
%
\begin{figure*}[!t]
\centering
\includegraphics[width=0.90\textwidth]{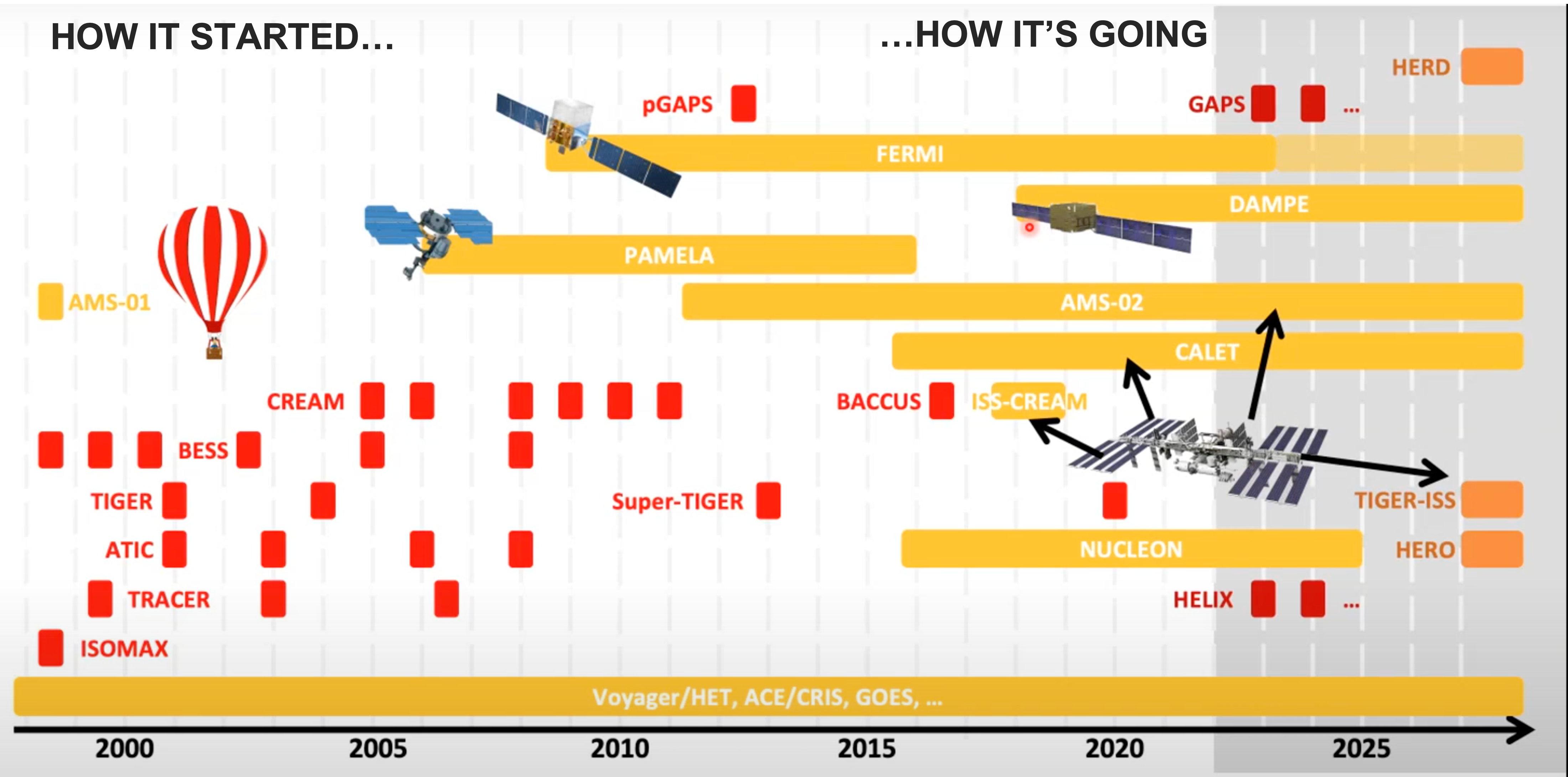}
\caption{\footnotesize%
  How it started\dots\, how it's going. A timeline of the accomplished, ongoing, and planned experiments of direct
  CR detection in space (yellow) or on stratospheric balloons (red) in the 3rd millennium. Courtesy of Alberto Oliva.
}
\label{Fig::ccCRMeasurementsMillennium}
\end{figure*}
%
Direct measurements on CRs at kinetic energies between multi-MeV/n to multi-TeV/n have been performed extensively, since the 60s,
by numerous experiments onboard stratospheric balloons, orbiting satellites, space probes or space stations.
The present paper is focused on the last two decades, when CR measurements from space gained an increased attention.
The main goals of these experiments have been the measurement of CR fluxes as function of the particle energy or rigidity,
along with the mass or charge composition, arrival direction, or temporal variations in connection with solar activity.
Prime goal of these experiments is the investigation of fundamental astrophysical mechanisms pertaining acceleration and transport and interactions
of cosmic particles in the interstellar or interplanetary space.
Other goals of the CR physics research are the detection for rare particles such as primordial antimatter nuclei,
strange quark matter or fractionally charged particles, along with the search of tiny signals of dark matter annihilation.
Experiment placed outside the atmosphere onboard space vehicles are particularly suitable
for the search of the rarest constituents of the cosmic radiation.

In the last two decades, several important results in CR physics have been achieved. 
Precise measurements up to the TeV scale have been conducted by
the magnetic spectrometers PAMELA \citep{Adriani2014Report} and AMS-02 \citep{Aguilar2021Report}.
The PAMELA experiment, \emph{Payload for Antimatter-Matter Exploration and Light-nuclei Astrophysics}, was launched on June 15th, 2006 from the Baikonur Cosmodrome, Kazakhstan,
and operated onboard the Russian satellite Resurs-DK1 till 2016.
The AMS-02 experiment, second and final version of the \emph{Alpha Magnetic Spectrometer} project, was installed on the International Space Station
on 19 May 2011 and will be operating for the station lifetime. 
Both experiments are designed to study high-energy CRs with focus on antiparticles and rare components.

Alternative to magnetic spectrometry, experiments based on calorimetric techniques are widely employed. 
Calorimeters measure the total CR energies by evaluated the energy deposition of the incoming particle and its associated shower.
Recent examples are the space born calorimeters are DAMPE, CALET, NUCLEON and ISS-CREAM, all launched in the past
few years \citep{Boezio2020,PengYuan2019,PicozzaMarcelli2019,Marocchesi2017}.
Along with space missions, experiments on high-altitude balloons are still playing an important role in the direct detection of CRs.
Recent examples are the flight campaigns of BESS, CREAM, ATIC, Super-Tiger, Baccus,
along with forthcoming experiments such as HELIX and GAPS that are expected to run in short.
A timeline of the recent and future projects of direct CR detection is illustrated in Fig.\,\ref{Fig::ccCRMeasurementsMillennium}.
In light of the several ongoing and planned experiments, the present epoch can be rightfully considered as a \emph{golden age} of CR measurements.

\section{Recent results} 
\label{Sec::Results}     

New-generation experiments of direct CR detection are probing the fine structure of the CR phenomenology and bringing important results.
In this recent \emph{golden age} of CR measurements, several unexpected features have been discovered in the CR energy spectrum,
and thus the study of CR acceleration and propagation processes has become more important than ever.
In the next paragraphs I will give an overview of the recent results on the CR direct measurements,
with a focus on the unexplained anomalies found in the CR spectrum, along with a discussion on their possible interpretations.

\subsection{High-energy protons and helium} 

In the last decade, the accumulation of several measurements on CR proton and helium with unprecedent precision
led to the fall of the
traditional view that a simple and universal power-law can represent the CR flux without deviations up to PeV energies.
A first challenge to the standard picture was provided by the observation of a discrepant hardening
in the protons and helium spectra. Several experiments such as CREAM, ATIC, PAMELA, and AMS-02  
measured a change of spectral slopes at kinetic energy $E \gtrsim 200$\,GeV/nucleon \cite{Adriani2014Report,Yoon:2011aa,Panov:2011ak,Aguilar:2015P,Aguilar:2015He,Aguilar2021Report}.
In addition, these experiments observed a tiny difference between the high-rigidity spectrum of CR protons and than that of helium or other nuclei
leading, in particular, to a remarkable decrease of the  p/He ratio at increasing rigidity: $p/He\propto{R}^{-\alpha}$, with $\alpha\cong{0.1}$.
These features cannot be explained within the conventional models of linear diffusive shock acceleration of CRs followed by
diffusive propagation in the ISM \cite{Kachelriess2019}.
Several works suggested to account for the
contribution of different accelerating sources, nonlinear effects in their propagation, or spatial-dependent diffusion
\cite{Kachelriess2015SNR,TomassettiDonato2015,Blasi:2012yr,Tomassetti2015PHeAnomaly,Tomassetti2015TwoHalo,Tomassetti2012TwoHalo}
%
\begin{figure}[!t]
\centering
\includegraphics[width=0.45\textwidth]{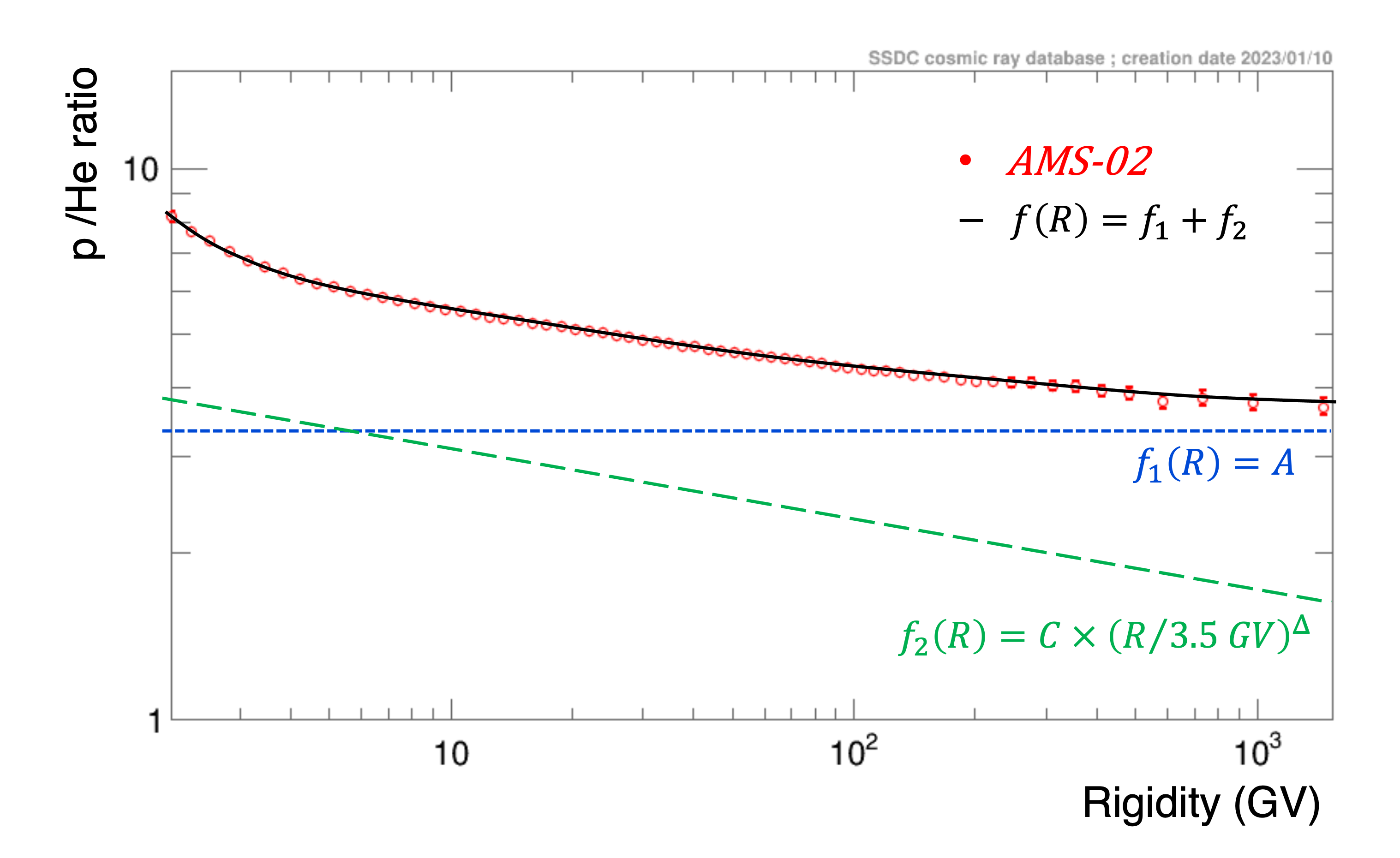}
\caption{\footnotesize%
  The p/He ratio as function of rigidity measured by AMS-02 between 2 GV and 2 TV.
  The ratio is steadly decreasing with increasing rigidity, as reported by recent data;
  but at TV rigidity, the AMS-20 data show that the ratio flattens \citep{Aguilar2021Report}.
}
\label{Fig::ccPHeRatioAMS02}
\end{figure}

On the other hand, such features may offer a clue to the origin of high-energy CRs. 
The origin of the spectral hardening can be clarified with high-energy measurements of secondary fluxes and secondary/primary ratios, see Sec.\,\ref{Sec::SecPriRatio}.
Regarding the decrease observed in the p/He ratio at increasing rigidity, recent observations by AMS-02 show that the ratio becomes flatter at TeV energies.
The measurement of the p/He ratio is shown in Fig.\,\ref{Fig::ccPHeRatioAMS02}. As seen, the data are well fitted with the sum of two components.
The soft component, more pronounced at low rigidities, is a power-law function of the type $f_{2}(R)=C(R/R_{0})^{\Delta}$, with $R_{0}=3.5$\,GV, $\Delta=-0.3$, and $C=3.30$.
At high rigidities, the hard component $f_{1}(R)=A$ dominates and the proton-to-helium flux ratio, with gradually approaches a constant value of  $A=3.15$ \citep{Aguilar2021Report}. 
An physical interpretation of this results is that a two-component scenario, where the total CR flux is made by two main classes of sources
with different p-He composition (\ie, different metallicity) and different spectral slopes (\eg, different ages) \citep{Tomassetti2015PHeAnomaly}.
%
\begin{figure}[!t]
\centering
\includegraphics[width=0.47\textwidth]{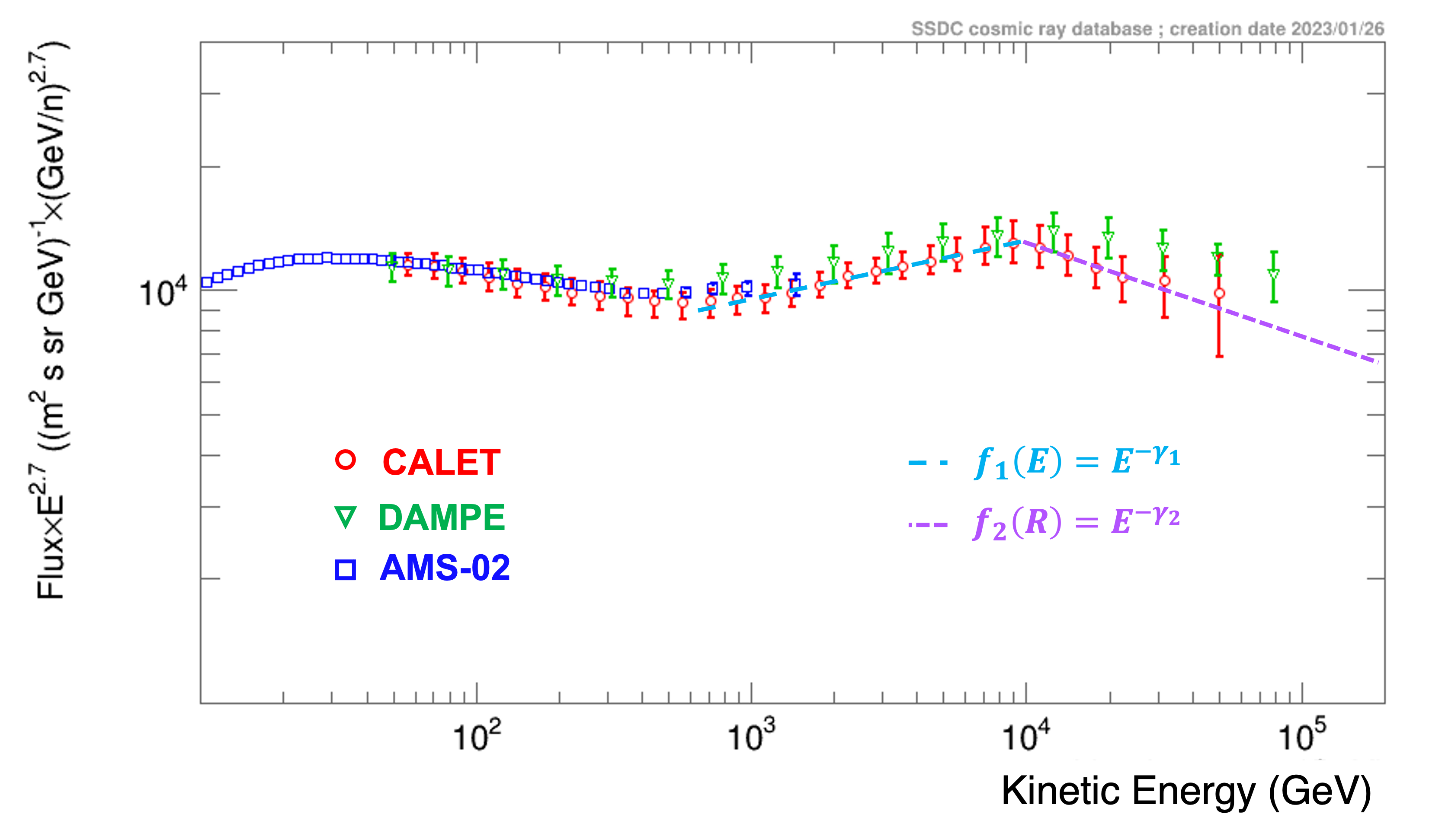}
\caption{\footnotesize%
  The proton flux as function of kinetic energy $E$  measured by CALET, DAMPE, and AMS-02 \citep{An2019,Adriani2022,Aguilar2021Report}.
  The flux is multiplied by $E^{2.7}$.
  The dashed lines illustrate the 10\,TeV break where the spectral index is found to change from $\gamma_{1}\approx{2.6}$ to $\gamma_{2}\approx{2.9}$.
}
\label{Fig::ccPHeRatioAMS02}
\end{figure}

More recently, another interesting and unexpected feature was observed from calorimeters DAMPE, CALET, NUCLEON and ISS-CREAM.
It consists in the observation a new “knee”, \ie, a spectral steepening, in the flux of CRs at about 10\,TeV of energy \citep{An2019,Alemanno2021,Adriani2022,Atkin2018,Choi2022}.
Such a features has been observed in both protons and helium fluxes. The spectral index of CR protons (helium) is found to change from 2.60 (2.5) to nearly 2.9 (2.8).
These result seems to suggest that the Galactic flux CRs arises from a mixture of different classes of sources, each characterized by distinct acceleration spectrum and cutoff.
After the CR spectral hardening at $\sim$\,100\,GeV, the break at GeV energies, the first and second knees at the multi-PeV scale, the EeV ankle, 
such a 10\,TeV softening is only a new piece of evidence against the old idea that the energy spectrum of CRs can be described by a simple and universal power-law.

\subsection{Primary and secondary nuclei} 
\label{Sec::SecPriRatio}                  

\begin{figure}[!t]
\centering
\includegraphics[width=0.47\textwidth]{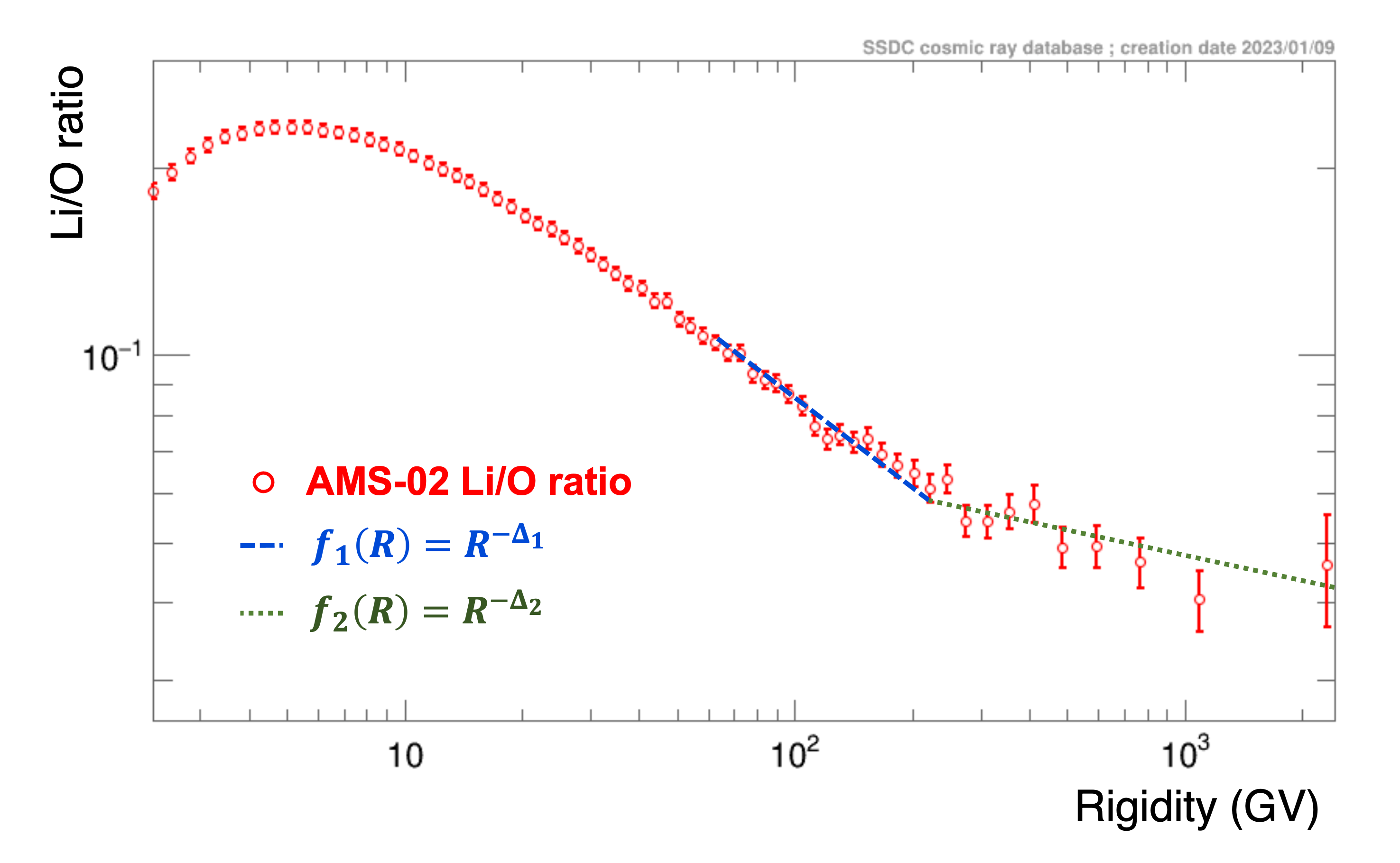}
\caption{\footnotesize%
  The Li/O ratio as function of rigidity measured by AMS-02 between 2 GV and 4 TV.
  At $R\gtrsim$\,60\,GV and up to $200$\,GV the ratio is steadily decreasing as power-law $\propto{R}^{-\Delta}$ with from $\Delta\approx{0.45}$,
  but at rigidities higher than $200$\,GV it shows a remarkable hardening to $\Delta\approx{0.25}$ \citep{Aguilar2021Report}.
}
\label{Fig::ccLiORatioAMS02}
\end{figure}

In the recent years, important results have been obtained with the precision measurement of the nuclear component in CRs.
The precise measurements of high-energy carbon and oxygen made by AMS-02 and CALET gives the important indication that
the CR spectral hardening is a common feature of all primary nuclei \citep{Aguilar2017HeCO,Adriani2020CaletC}. 
A further insight about the origin of the hardening comes from the AMS-02 flux measurement of secondary Li-Be-B elements  from 2 GV to 4 TV of rigidity,
along with several secondary to primary ratios such as B/C, Be/C, Li/O, et cetera \citep{Aguilar2016BC,Aguilar2018LiBeB}.
In these ratios, a 200\,GV spectral hardening has been observed by AMS-02, revealing that the change of slope in the
energy spectra of secondary nuclei (Li, Be, B) is stronger than that of primary nuclei (C, O).
As an example, Fig.\,\ref{Fig::ccLiORatioAMS02} shows the Li/O ratio measured by AMS-02 as function of rigidity.
As seen from the figure, in the high-rigidity region ($R\gtrsim$\,60\,GV) the ratio can be described by a double power-law function.
The transition is seen to take place at the rigidity $R\approx\,192$\,GV, where the log-slope of the ratio changes
from $\Delta\approx{0.25}$ to $\Delta\approx{0.45}$.
%
%
%
%
%
More recently, such a hardening in secondary to primary ratios has been confirmed by
further measurements of the B/C ratio made by NUCLEON, DAMPE, and CALET \citep{Alemanno2022DampeBC,Adriani2022CaletBC}.
These results provides a strong clue for the confirmations for a \emph{diffusive} origin of the CR spectral hardening \citep{Tomassetti2012TwoHalo,Blasi:2012yr}.
In fact, if the hardening originate from intrinsic properties of the CR accelerators, no features should be expected in the secondary to primary ratios.
On the other hand, change of slope in the secondary to primary ratios are expected
if the hardening arises from CR propagation effects in the ISM \citep{Tomassetti2012TwoHalo}.
In practice, an interstellar hardening can be produced by the self-generation of turbulence by CRs \citep{Blasi:2012yr}, by a non-separable diffusion coefficient
in the energy and space coordinates \citep{Tomassetti2012TwoHalo,Tomassetti2015TwoHalo}, or by a combination of the two effects \citep{Evoli2018}.

\subsection{Intermediate and high-charge nuclei} 
\label{Sec::HeavyNuclei}                         

\begin{figure}[!t]
\centering
\includegraphics[width=0.45\textwidth]{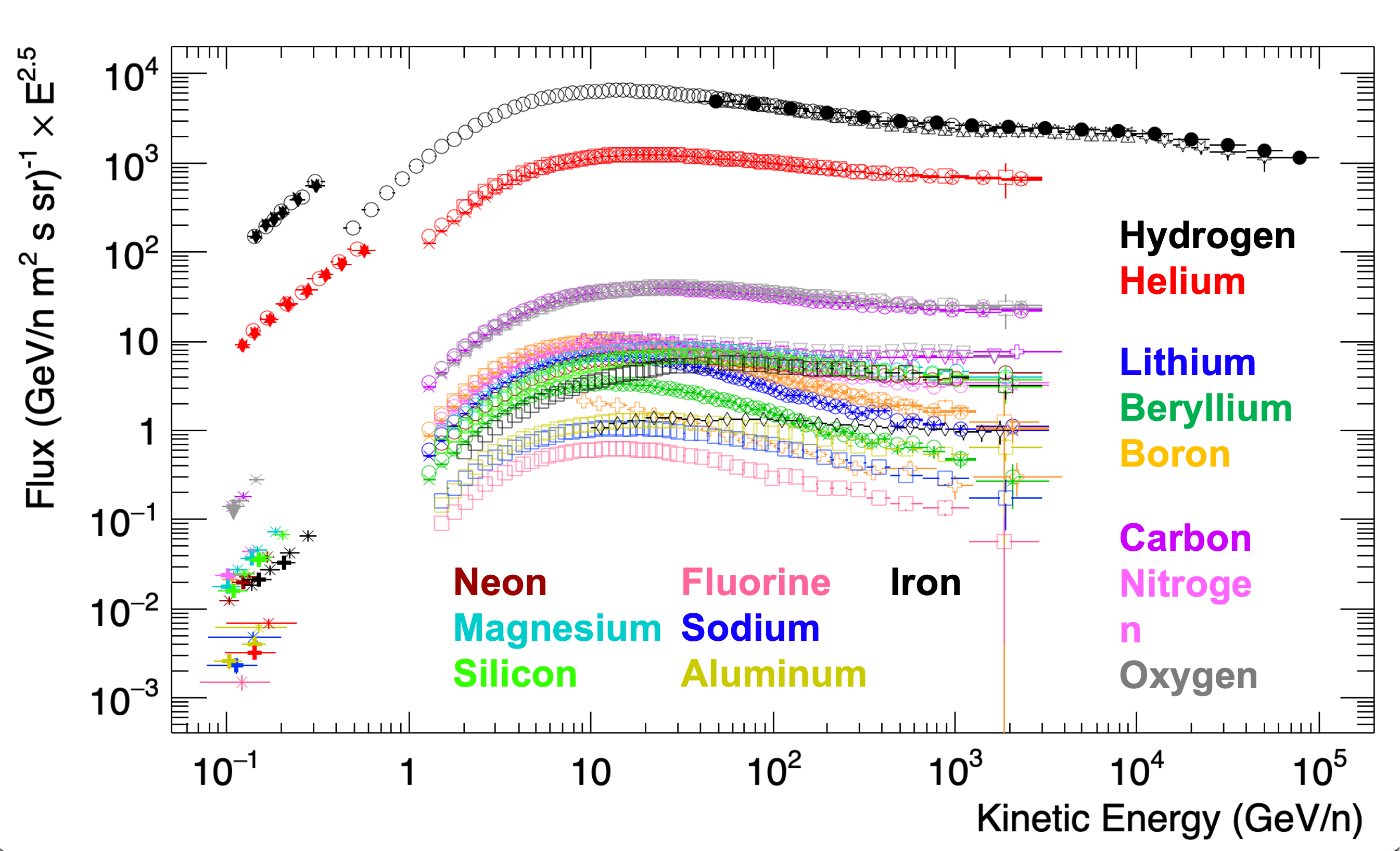}
\caption{\footnotesize%
  Compilation of flux measurements for many elements between He and Fe multiplied by $E^{2.5}$. The data shown here comes from Voyager-1 (sub-GeV), AMS-02 (Gev to TeV), CALET and DAMPE (multi-GeV to multi-TeV).
}
\label{Fig::ccAllNucleiAMS}
\end{figure}

With the recent measurements of several nuclear species at higher charges, from $Z=9$ to $Z=28$, new intriguing anomalies are emerging.
A compilation of measurements for several nuclear species is shown in Fig.\,\ref{Fig::ccAllNucleiAMS}, where the flux of CR elements
from H to Fe are plotted as function of energy and multiplied by $E^{2.5}$.
In the following I will recapitulate some of the most interesting and puzzling features.
%

\begin{figure}[!t]
\centering
\includegraphics[width=0.45\textwidth]{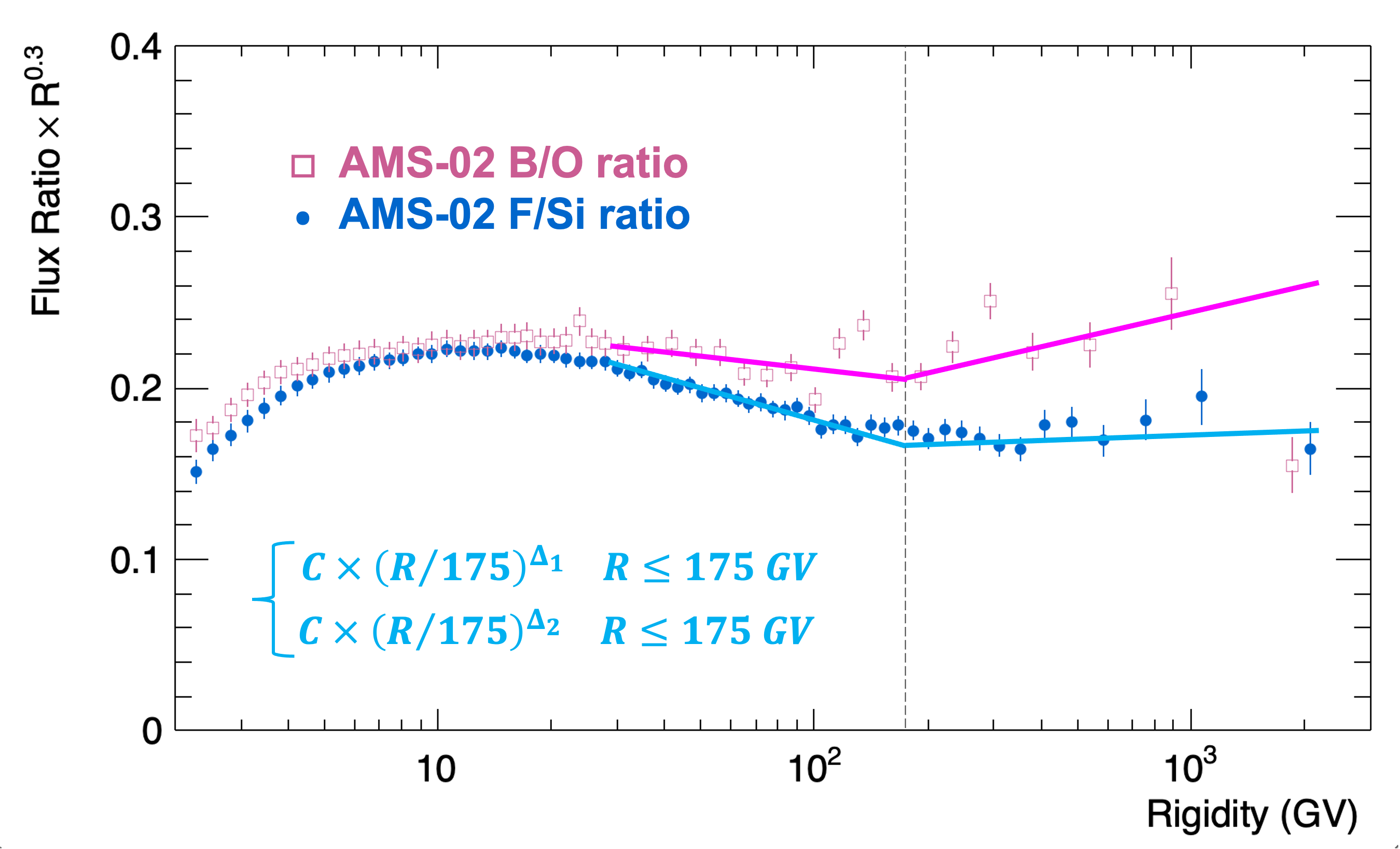}
\caption{\footnotesize%
  Measurements made by AMS-02 of the F/Si ratio as function of rigidity compared with the B/O ratio rescaled by 0.39. Both ratios are multiplied by $R^{0.3}$. The ratios are described with a double power-law functions with break rigidity of 175\,GV \citep{Aguilar2021F}.
}
\label{Fig::ccFSiRatio}
\end{figure}

An important measurement recently made by AMS-02 is the fluorine flux ($Z=9$) \citep{Aguilar2021F}.
The experimental detection of fluorine in CRs is challenging because it is under-abundant in comparison
with nuclei of close atomic number such as oxygen ($Z=8$) or neon ($Z=10$).
For this reason, the past measurements on the fluorine flux are scarce and affected by sizeable errors. 
In the cosmic radiation, fluorine is a secondary element \ie, it is entirely produced by collisions of
heavier nuclei such as Ne, Mg, Si, or Fe with the interstellar gas.
The rigidity dependence of the fluorine flux should then be similar to that of boron.
In particular, the F/Si ratio is expected to decrease with
increasing rigidity like the B/O ratio or other secondary to primary ratios.
The precise measurements made by AMS-02 between 2 GV and 2 TV show that the F/Si ratio subjected to a high-energy hardening, similarly to that observed in other ratios.
In particular, at rigidity above $R\approx$\,175\,GV, a change of slope of 0.15 is reported, consistently with the hardening of the B/O ratio.
However, in the whole rigidity range, the decrease of the F/Si ratio is found by AMS-02 to be shallower than that of the B/O ratio.
In other words, the ratio between F/Si and B/O decreases with rigidity.
The two ratios are compared in Fig.\,\ref{Fig::ccFSiRatio}.
The different rigidity dependence of the F/Si ratio is not captured by the existing models of CR propagation and the origin of such a different is currently unexplained.
%

Along with fluorine, recent measurements from AMS-02 involve the high-energy fluxes of Ne, Mg, and Si nuclei \citep{Aguilar2020NeMgSi}. 
These nuclei are mostly of primary origin, similarly to H, He, C, and O nuclei.
However, in the high rigidity region ($R\gtrsim\,80\,$GV), the rigidity spectra of Ne-Mg-Si elements are found to be softer than those of other primaries such as He, C, and O.
Although both groups must be of primary origin, the observed spectral difference may suggest  that different CR sources contribute to the flux of different elements.
In addition, other recent measurements by AMS-02 seems to indicate that Sulfur energy spectrum belong to the Ne-Mg-Si class of intermediate nuclei.
These results suggest that some of the traditional assumptions of the standard CR propagation models are not longer valid.
In particular, the assumption that the injection of CRs in the ISM can be described by a simple universal and composition-blind acceleration function
appears no longer supported by the AMS-02 nuclear data \citep{Korsmeier2022}.

Other measurements made by AMS-02 involve mixed nuclei such as nitrogen, in particular, which is a mix of primary and secondary particles \citep{Aguilar2018N}.
Nitrogen, similarly to elements such as aluminum and sodium, is accelerated in primary sources of CRs. However the
abundance of nitrogen generated by fragmentation of heavier species is comparable to its primary component accelerated in supernova remnants.
The nitrogen flux is a 50-50 mix of primary and secondary components having distinct spectral properties. Since the primary components are generally harder than the secondary components,
one might expect that the former dominate the total flux at higher rigidity while the latter dominate at low rigidity.
The recent measurements from AMS-02 have shown that, below 100 GV, the N flux and its spectral index are similar to that of Na,
while above 100 GV the N flux becomes similar to the flux of Al \citep{Aguilar2021NaAlN}. The origin of such a behavior is not understood.

\begin{figure}[!t]
\centering
\includegraphics[width=0.465\textwidth]{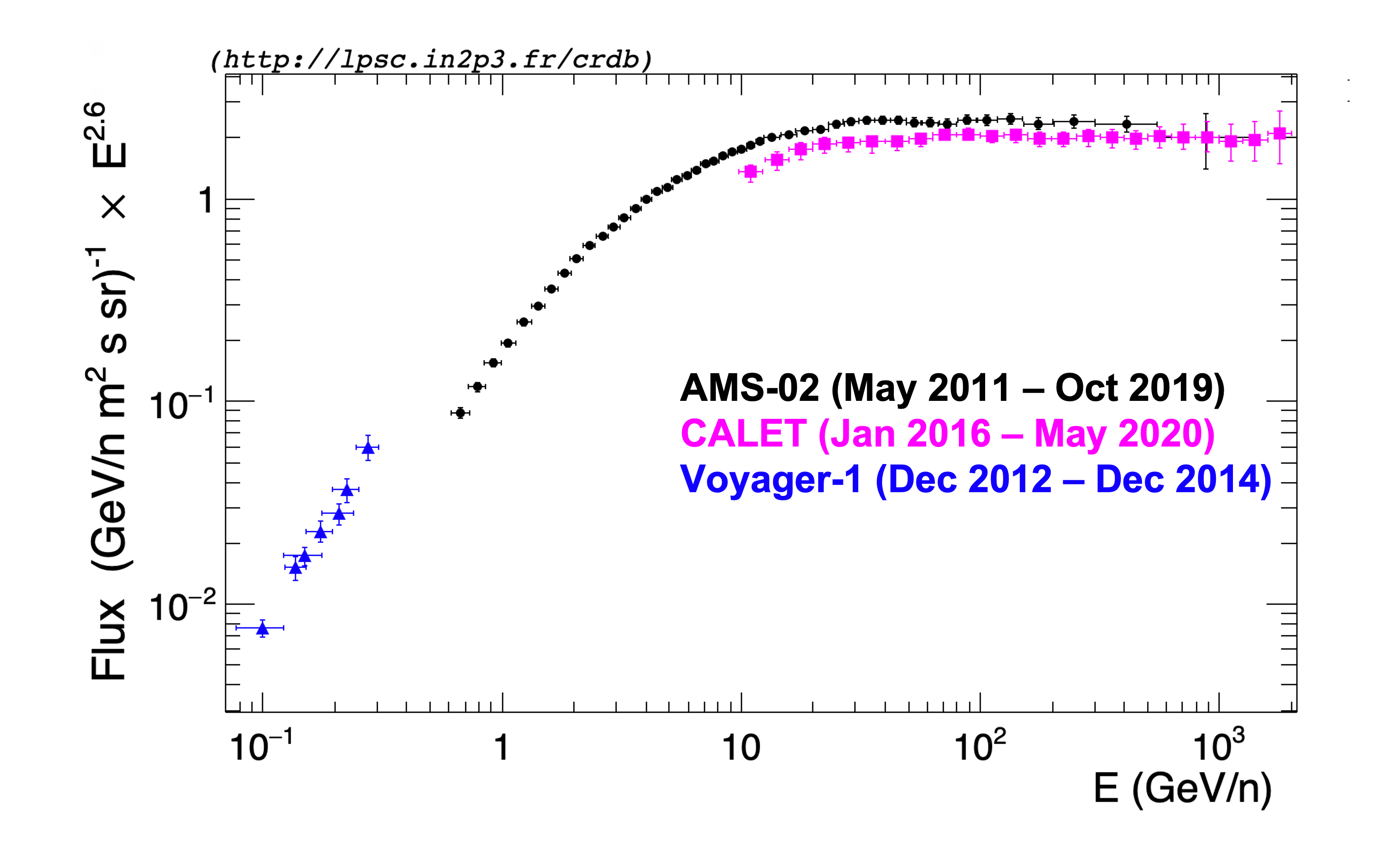}
\caption{\footnotesize%
  Flux of CR iron nuclei as function of kinetic energy per nucleon $E$ measured by CALET, AMS-02, and Voyager-1 \citep{Aguilar2021Fe,Adriani2021CaletFe,Cummings2016}.
  The flux is multiplied by $E^{2.6}$.
}
\label{Fig::ccIronFlux}
\end{figure}

The high-energy flux of CR iron has recently been measured by AMS-02, CALET, and NUCLEON \citep{Aguilar2021Fe,Adriani2021CaletFe,Grebenyuk2019NucleonFe}.
These data cover a large dynamic range, from below one GeV to a few TeV of kinetic energy per nucleon. The CR iron flux data are shown in Fig.\,\ref{Fig::ccIronFlux}. 
%
%
Contrary to lighter species, it can be seen from the figure that the flux of CR iron does not show a prominent spectral hardening. 
However, one may expect that the spectral hardening is less pronounced for heavier elements, due to the competing action of the destruction processes
which hardens the low-energy part of the spectrum. In fact, the cross-section for CR destruction in the ISM increases with the nucleus mass.
Nonetheless, the new measurements of the iron spectrum appears in contrast with the calculations of current models of CR transport, and in contradiction with other low-energy data
from past experiments such as HEAO, ACE-CRIS and Voyager \citep{Schroer2021}. 
There are various possible explanations for the iron anomaly: the different injection mechanism, the presence of multiple breaks in the acceleration spectrum,
a mis-estimation of the iron fragmentation cross-sections, unaccounted effects in solar modulation.
In a recent work, the analysis of combined measurements on CR iron revealed
that the low-energy discrepancy between the AMS-02 and earlier experiments ACE/CRIS and Voyager-1
can be explained by the presence of a small bump in Fe spectrum \citep{Boschini2021}.
Such a bump, appearing between 1 and 2 GV of rigidity, would originate from a local accelerator of low-energy CRs. 
There are various indications for the existence of a local sources, \eg, from the observed excess of  $^{60}$Fe isotopes
in deep-ocean samples or in the lunar regolith, or from the models of the Local Bubble formation from past supernova activity. 
A 2-3 Myr old supernova occurred within 100 pc from Earth would explain various features observed in primary and secondary CR species \citep{Kachelriess2015SNR,Kachelriess2018}.
Besides the iron anomaly, similar features may be present in the flux of aluminum and fluorine,
although these species  are observed to behave differently to other primary elements such as He, C, or Si nuclei  \citep{Boschini2022}.
Since the low-energy region the CR flux is significantly reshaped by solar modulation, nuclear interactions, ionization and Coulomb scattering, a better investigation is needed.

Regarding the low-energy physics phenomena, and in particular the heliospheric modulation of CRs, important observational advance and theoretical understanding
have been gained in the recent decades.
The study of solar modulation has been limited for long time by the scarcity of long-term CR data on different species,
and by the poor knowledge of their local interstellar spectra. The entrance of Voyager-1 in interstellar space, in 2012 provided us
with the very data on the CR fluxes unperturbed by solar activity. Moreover, the realization of long-term experiments in space
such as AMS-02, PAMELA, or CALET, along with spacecraft experiments, provided us with a large wealth of time-resolved and multichannel data
on CR particles, nuclei, and antimatter \citep{Cummings2016,Israel2018CRIS}.

\subsection{Cosmic leptons} 
\label{Sec::CRLeptons}      

\begin{figure}[!t]
\centering
\includegraphics[width=0.465\textwidth]{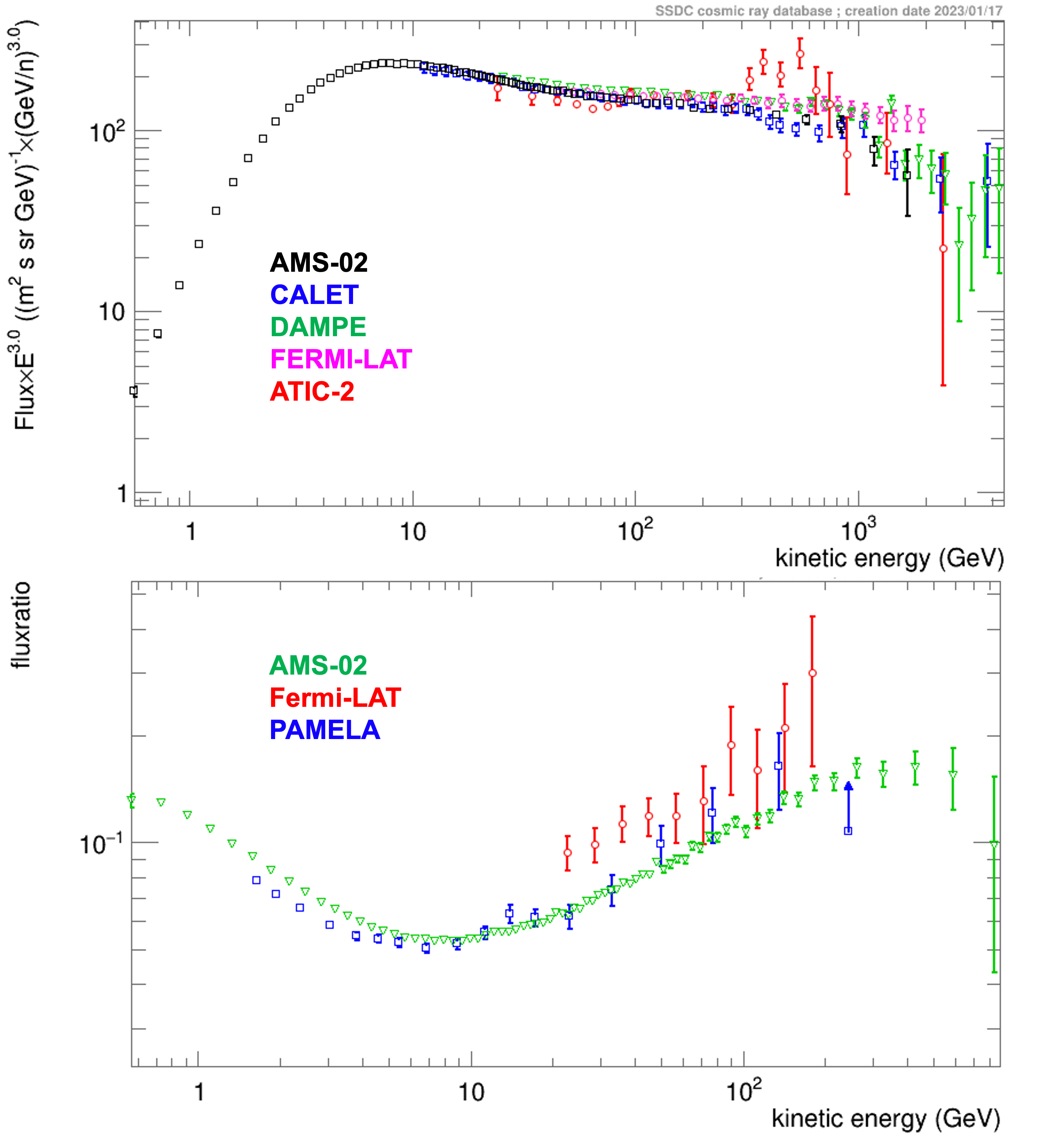}
\caption{\footnotesize%
  Compilation of recent measurements on light leptons. Top: the all-electron flux $(e^{+} + e^{-})$ as function of the CR energy $E$ multiplied by $E^{3}$.
  Bottom: the positron fraction $e^{+}/(e^{+} + e^{-})$ as function of energy \citep{Adriani2014Report,Aguilar2021Report}.
}
\label{Fig::ccCRLeptons}
\end{figure}

I will now turn the attention to the leptonic component of the cosmic radiation, namely electrons and positrons.
Several experiments measure the so-called \emph{all-electron} spectrum $e^{+} + e^{-}$, \ie, the total flux resulting from the sum of electrons and positrons.
A compilation of all-electron flux data is shown in Fig.\,\ref{Fig::ccCRLeptons}. From the figure, some interesting features can be observed.
In the multi-TeV region, the recent data from CALET and DAMPE reported a significant spectral softening at about 1 TeV,
confirming previous indications that were coming from indirect measurements \citep{Ambrosi2017Dampe,Adriani2018Dampe}.
The origin of such a feature is debated. It may be the result of a local  accelerator with maximum energy in the TeV scale,
or it may be related to the radiative losses of CR leptons during their propagation.
Magnetic spectrometers have also separated the two electron and positron individual components.  
High-energy data in the CR positron flux are of considerable interest.
In traditional models of CR propagations, cosmic positrons are of secondary origin and are generated by
collisions of high-energy protons with the gas.
%
%
%
%
In a conventional picture of CR propagation where all positrons are of secondary origin,
the energy spectrum of CR positrons is expected to decrease more rapidly to that of electrons.
Such a simple expectation is in clear contrast with observations of the so-called \emph{positron fraction} $e^{+}/(e^{+} + e^{-})$, \ie,
the ratio between the positron flux to the sum of positron and electron fluxes.

Some recent measurements on the positron fraction from PAMELA, Fermi-LAT, and AMS-02 are shown in Fig.\,\ref{Fig::ccCRLeptons},
where interesting features can be noticed. First, the low-energy data ($E\lesssim\,5$\,GeV) are approximately consistent
with the standard models of secondary production where CR positrons are generated by collisions of CRs with the gas \citep{Adriani2014Report,Aguilar2021Report}.
However, the low-energy region is highly sensitive to charge-sign dependent effects in the heliospheric CR modulation.
On the other hand, in the high-energy region ($E\gtrsim$\,10\,GeV), all data show a significant \emph{excess} of CR positrons, with respect to conventional expectation.
Instead of decreasing like secondary-to-primary ratios, the positron fraction increases steadily up to $\sim$\,200 GeV.
The search for the origin of high-energy positrons have attracted widespread interest of the CR physics community.
The proposed explanations for the \emph{positron excess} include dark matter particle annihilation into $e^{\pm}$ pairs,
or nearby astrophysical sources of primary positrons such as pulsars or supernova remnants.
In particular, the most accepted explanation is the production of high-energy pairs $e^{\pm}$  in the magnetosphere
of nearby pulsars such as Geminga or PSR B0656+14.
The issue however is still debated. To clarify on the origin of the anomaly, high-energy data on CR positrons are essential.
From the AMS-02 data, it can also be noticed that the positron fraction peaks at the energy scale of about 500\,GeV.
If the positron excess is caused by a nearby accelerator, the observation of a peak gives precious indications on its properties.
In the dark matter annihilation hypothesis, the peak would be directly related to the mass of the hypothetical dark particles, which,
from the AMS-02 data, would lie around the TeV scale.
To explain the positron data in terms of annihilation of dark matter particles, however, requires a unrealistically
high value for their annihilation rate: $\langle\sigma_{A}v\rangle\gtrsim\,10^{24}$\,cm$^{3}$/s. 
For comparison, the value required for the thermal production of dark matter in the early universe is $\langle\sigma_{A}v\rangle\sim\,2{\cdot}10^{26}$\,cm$^{3}$/s.
Another unsolved problem with the dark matter interpretation is how to conciliate the CR positrons data with the high-energy antiproton data.

\subsection{Antiprotons and antinuclei} 
\label{Sec::Antiprotons}                

Similarly to positrons, antiprotons or light antinuclei are expected to be generated by secondary
production processes, that is, in nucleus-nucleus collisions between CRs and interstellar gas.
However, the subsequent processes of diffusive propagation and interactions in the Galaxy are similar to those
of other hadronic or nuclear components of CRs such as protons, or other light nuclei. 
In particular, in the high-energy limit, the antiproton to proton ratio $\bar{p}/p$ must show a similar energy dependence of the B/C ratio,
\ie, the flux of secondary antiprotons should be steeper than the proton flux.
Moreover, nuclear antimatter cannot be produced in the magnetosphere of pulsars, which is the dominant astrophysical explanation for the positron excess.
For this reasons, antiprotons can be considered as a clean and powerful probe
for the search of exotic mechanisms of CR antimatter production such as dark matter or evaporation of primordial black holes.
%
%
\begin{figure}[!t]
\centering
\includegraphics[width=0.465\textwidth]{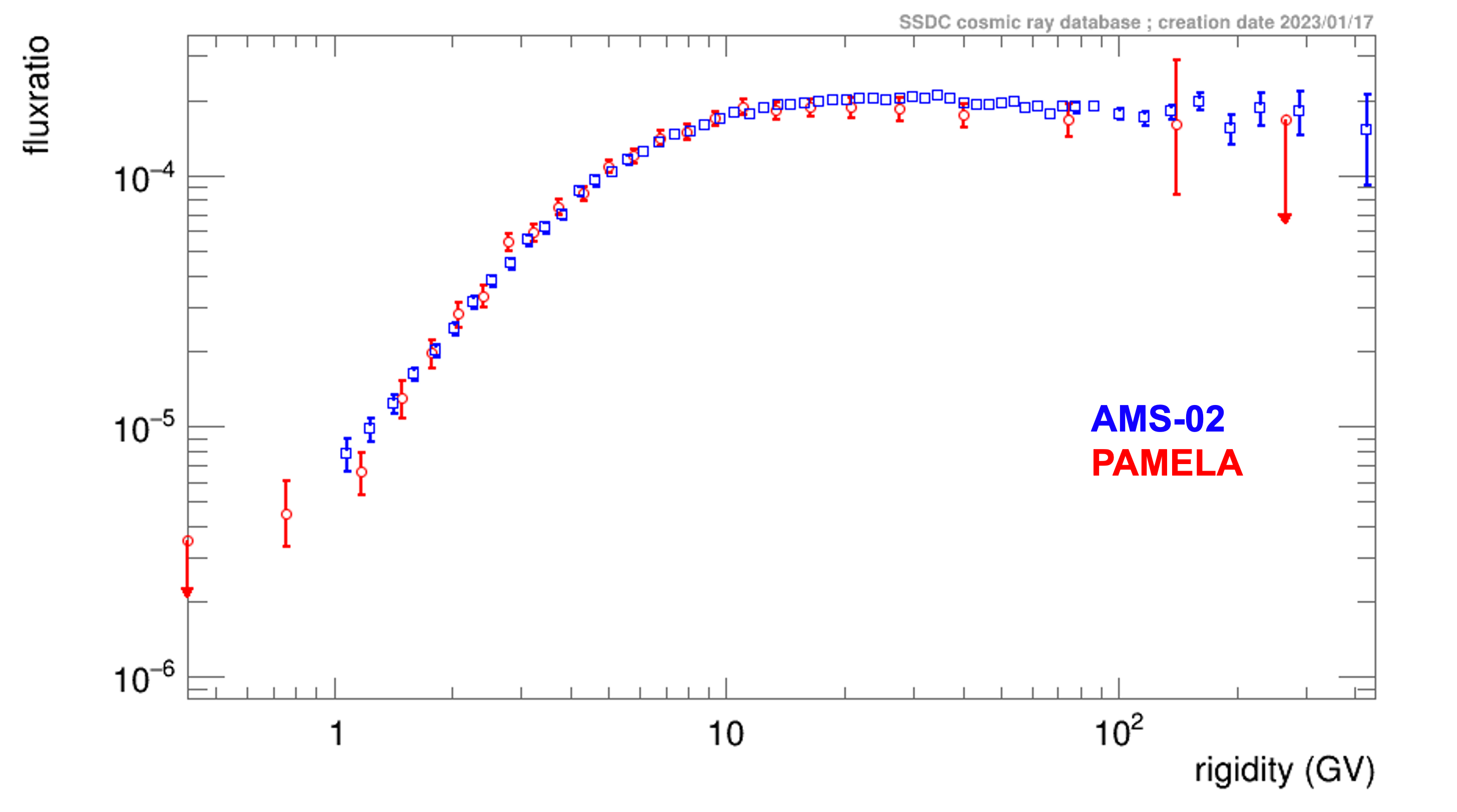}
\caption{\footnotesize%
Antiproton to proton ratio as function of rigidity measured by AMS-02 and PAMELA \citep{Adriani2014Report,Aguilar2021Report}.
}
\label{Fig::ccPbarPRatio}
\end{figure}
%
%
Recent measurements of the antiproton to proton ratio from AMS-02 and PAMELA are shown in Fig.\,\ref{Fig::ccPbarPRatio}.
In the figure, some interesting features can be noticed. At low rigidities, ($R\lesssim\,10$\,GV)  fall-off of the ratio is approximately consistent
with the kinematics for the antiproton production. The dominant production channel is $p+p\rightarrow \bar{p}+3p$, where high-energy protons of the
cosmic radiation collide with ISM nucleons at rest.
At high rigidities ($R\gg\,10$\,GV), the $\bar{p}/p$ ratio appears remarkably energy independent,
flatter than that expected by previous calculations. Such a behavior is not easy to explain in terms of standard production of secondary antiprotons.
However, the data do not present prominent features expected from dark matter annihilation models such as peaks or bumps. 
Using these data, one can place strong limits to the rate of dark matter annihilation into quark-antiquark pairs. 

On the other hand, with the high-precision nuclear data on secondary to primary ratios, \eg, on the B/C ratio,
one can set tight constraints on the parameters of the \emph{astrophysical background} from the secondary production mechanism.
Using state of the art models of CR propagations, some authors have claimed indications of an antiproton excess in the $\sim$\,5-20 GV
region with respect to secondary production \citep{Cuoco2019,Cui2017}.
According to these authors, such an excess has a 3-sigma significance after accounting for theoretical and experimental uncertainties.
Several other studies, however ,have claimed that the antiproton data are consistent with secondary production mechanism
once all sources of uncertainties are properly accounted \citep{Evoli2015,Feng2016,Boudaud2020,Calore2022}.
Although the significance for such an antiproton excess is debated, 
it is interesting to notices that the fits to the $\bar{p}/p$ data in terms of dark matter models are in nice agreement with those
accounting for the so-called \emph{GeV excess} in the Galactic center observed with gamma rays \citep{Cuoco2019}.
Moreover, the resulting annihilation rate for dark matter particles into quark-antiquark pairs that generate antiprotons
is of the order of $\langle\sigma_{A}v\rangle\sim\,10^{26}$.
This value is remarkably well in agreement with cosmological estimates of the thermal production of dark matter in the early universe.

While the antiproton data have generated a renewed interest in the search of dark matter using cosmic antimatter,
a conclusive and unambiguous answer might come with the detection of antideuteron and antihelium.
These species,  in fact, the astrophysical background appears reduced in comparison of the expected signal of
dark matter annihilation.
At the time of this review, no antinuclei have been reported in the cosmic radiation.
The best upper limits have been set from antimatter spectrometers PAMELA and BESS \citep{Beatty2022,Boezio2020}.
Release of new data or upper limits is await from the AMS-02 experiment. After nearly 12 years of observations,
AMS-02 should have reached a high level of sensitivity for testing several existing dark matter models.

\section{Future missions}      
\label{Sec::FutureExperiments} 

In this section, I will overview some of the forthcoming or proposed missions for direct detection of CRs.
%
%
\begin{figure}[!t]
\centering
\includegraphics[width=0.465\textwidth]{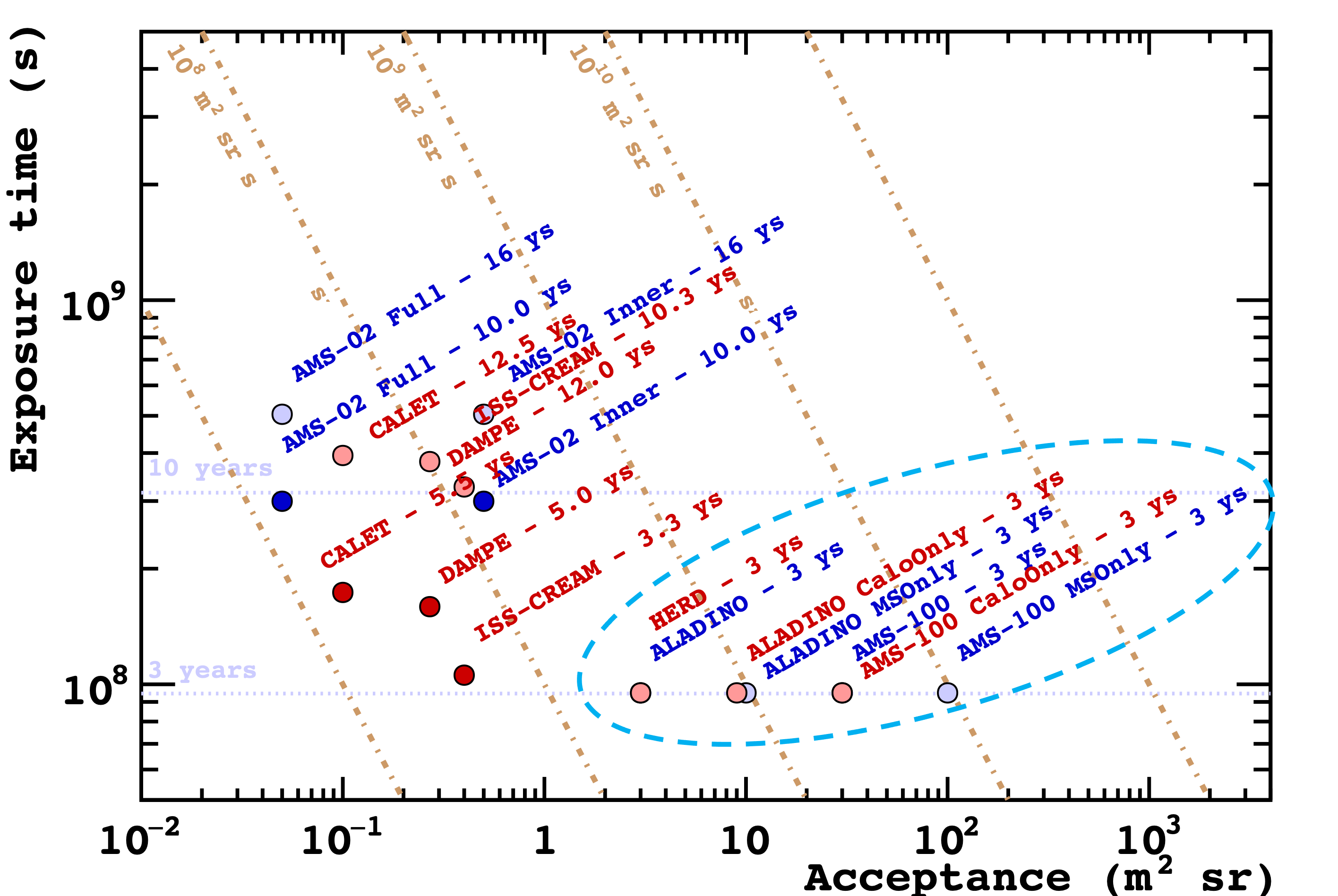}
\caption{\footnotesize%
  Diagram of present and future experiments in the 2D plane exposure time vs acceptance. The experiments are placed in various
  configurations of detection setup \citep{Adriani2022ALADInO}. 
  For proposed missions such as HERD, ALADInO and AMS-100, an observation time of 3 years is set as benchmark. Courtesy of Matteo Duranti.
} 
\label{Fig::ccFutureExperiments}
\end{figure}

One key project is the General AntiParticle Spectrometer (GAPS), which is set to begin its first flight campaign in the coming months.
The GAPS instrument is a novel concept of nuclear antimatter spectrometer. 
Its detection strategy of GAPS is based on the formation and decay of exotic atoms that follow the absorption of antinuclei in the active material \citep{Aramaki2016GAPS}. 
The experiment has planned at least two balloon flights in Antarctic. According to its projected performance, with a cumulate exposure
of $\sim$\,100 days of observation GAPS would reach a sensitivity  of ${\sim}10^{-6}$\,nuc$\,m^{-2}s^{-1}sr^{-1}GeV^{-1}$ for antideuterons of kinetic energy below $250$\,MeV/n.

Another balloon-borne project is the HELIX experiment, which aims to measure the chemical and isotopic abundances of light CR nuclei. 
HELIX consists of a 1\,T superconducting magnet with a high-resolution drift-chamber tracker, time of flight detector, and a ring-imaging Cherenkov detector \citep{Park2022HELIX}.
Measurements by HELIX, especially of $^{10}$Be from 0.2 GeV/n to beyond 3 GeV/n may provide an essential set of data for the study of CR propagation.
The experiment is planned to have a long-duration balloon flight during an upcoming balloon campaigns in the northern or southern hemisphere. 

A proposed space-based is the Penetrating Particle Analyzer (PAN): a modular magnetic spectrometer conceived to precisely
measure the CR flux and composition from 100 MeV/n to 20 GeV/n, \ie, at relatively low energies.
The PAN is lightweight and low-power, making it suitable for deep space and interplanetary missions  \citep{Wu2019PAN}.
It may study Galactic CRs, solar and trapped particles, as well as being used as a radiation monitor for space weather.

The High Energy cosmic Radiation Detection (HERD) project is the next \emph{big science} space-based calorimenter.
HERD is set to be a payload on the China Space Station.
HERD aims to make high-precision, high-statistics spectral measurements of CR nuclei up to PeV energies and will also serve as an observatory
for $\gamma$-rays at energy between few hundred of MeV up to 1 TeV, contributing to multi-messenger astronomy \citep{Gargano2019HERD}.

Two major future missions for detecting antimatter in space are currently under study.
The Antimatter Large Acceptance Detector In Orbit (ALADInO) is a magnetic spectrometer designed to have a
nearly $4\pi$ field of view and a geometrical acceptance of about $10$\,m$^{2}$sr \citep{Battiston2021Aladino}.
Its silicon microstrip tracker system reaches a maximum detectable rigidity of 20 TV and allows for redundant particle identification capabilities. 
By design, the ALADInO tracking system is covered by a wheel of plastic scintillators and incorporates a 60 radiation length core calorimeter \citep{Battiston2021Aladino,Adriani2022ALADInO}.
The ``ultimate'' big science experiment for CR antimatter is represented by AMS-100 \citep{Shael2019AMS100}.
Conceived a science platform for astrophysics in the high-energy domain, AMS-100 is able to a achieve a geometric factor
of 100\,m$^{2}$sr and a maximum detectable rigidity of 100 TV. With a 70 radiation length calorimeter, AMS-100 is also sensitive to
PeV energies CR nuclei and high-energy $\gamma$-rays.

Both experiments are designed for having long-duration missions in deep space orbits.
Compared to existing experiments, these missions may significantly improve the sensitivity for observing
of new phenomena in CRs, particularly antimatter.
From both experiments, the detection secondary antideuterons is guaranteed from any plausible model of astrophysical background.
In Fig.\,\ref{Fig::ccFutureExperiments}, some of the present and future experiments are placed in a 2D diagram representing exposure time vs acceptance.
The experiments are placed in various configurations of detection setup associated to different acceptance \citep{Aguilar2021Report,Adriani2022ALADInO}.
For the proposed experiments such as HERD, ALADInO and AMS-100, an observation time of 3 years is set as benchmark.
\\

\section{Conclusions}     
\label{Sec::Conclusions}  

In conclusion, the field of CR physics has seen substantial progress in recent years,
with new-generation experiments of direct CR detection bringing important results
and opening a new era of precision physics.
Despite this progress, the challenge of fitting all the observations
into a comprehensive and agreeable picture for the origin and propagation of CRs remains.
The increase in accuracy and sensitivity in detection is also raising new questions
and challenges for interpretation of the data,
suggesting the need for a critical revision of the current paradigm.
Nevertheless, the proposed or planned projects of CR detection address important
science questions in multi-messenger astrophysics, and may lead to transformative
advances in the field of CR physics, giving us a deeper understanding of the Universe.

\section{Acknowledgements} 

I would like to express my gratitude to the conference organizers for inviting me to give this talk.
Most of the data and plots shown in this review are extracted from the \texttt{CosmicRays} web interface of SSDC-ASI \citep{DiFelice2017}
or from the \texttt{CRDB} databse of LSPC-Grenoble \citep{Maurin2020}.
I thank C. Evoli, A. Oliva, and M. Duranti for providing me with nice figures.
I acknowledge the support of Italian Space Agency under agreement ASI-UniPG 2019-2-HH.0
and the program Fondo Ricerca di Base 2021 of the University of Perugia, Italy.



\end{document}